\documentclass[10pt,twocolumn,superscriptaddress]{revtex4-1}
\usepackage[T1]{fontenc}
\usepackage{mathpazo}
\usepackage{graphics}
\usepackage{graphicx}
\usepackage{amsmath}
\usepackage{amssymb}
\usepackage{amsthm}
\usepackage{multido}
\usepackage{psfrag}
\usepackage{verbatim}
\usepackage{mathrsfs}
\usepackage{xcolor}
\usepackage{hyperref}
\usepackage{psfrag}
\usepackage{soul}
\usepackage{cancel}
\usepackage[normalem]{ulem}

\newcommand{\E}{{\mathcal E}}		%  Generic superoperator
\newcommand{\K}{\mathcal K}
\newcommand{\ad}{\mathrm{ad}}		% Commutator superoperator
\renewcommand{\L}{\mathcal{L}}		% Original Lindblad superoperator
\newcommand{\eff}{\mathrm{eff}}

\newcommand{\bra}[1]{\ensuremath{\langle#1|}}
\newcommand{\ket}[1]{\ensuremath{|#1\rangle}}

\renewcommand{\P}{\mathcal P}	% Projection superoperator
\newcommand{\PP}{\mathscr P}
\newcommand{\U}{\mathcal U}
\newcommand{\dg}{\dagger}
\newcommand{\Q}{\mathcal Q}	% Projection superoperator
\renewcommand{\O}{\mathcal O}
\newcommand{\tr}{\mathrm{tr}}	% trace
\newcommand{\id}{\mathcal{I}}
\newcommand{\coh}{\mathfrak C}%{\mathrm{Coh}}
\newcommand{\sfR}{\textsf{R}}

\newcommand{\sfC}{\textsf{C}}
\newcommand{\D}{\mathcal D}
\newcommand{\sfP}{\textsf{P}}
\newcommand{\sfD}{\textsf{D}}

\newcommand{\sfT}{\textsf{T}}
\newcommand{\diag}{\mathrm{diag}}
\renewcommand{\H}{\mathscr H}
\newcommand{\sfW}{\mathsf W}

\begin{document}
\title{Witnessing quantum coherence in the presence of noise}
\author{Alex Monras}
\affiliation{F\'isica Te\`orica: Informaci\'o i Fenomens Qu\`antics,
Universitat Aut\`onoma de Barcelona, ES-08193 Bellaterra (Barcelona), Spain}
\affiliation{Centre for Quantum Technologies, National University of Singapore, 3 Science Drive 2, Singapore 117543}
\author{Agata Ch\k{e}ci\'{n}ska}
\affiliation{Centre for Quantum Technologies, National University of Singapore, 3 Science Drive 2, Singapore 117543}
\author{Artur Ekert}
\affiliation{Centre for Quantum Technologies, National University of Singapore, 3 Science Drive 2, Singapore 117543}
\affiliation{	Mathematical Institute, University of Oxford, 24-29 St Giles', OX1 3LB, UK}
\begin{abstract}
	We address the problem of assessing the coherent character of physical evolution. We take the quantum Zeno effect (QZE) as a characteristic trait of quantum dynamics, and derive relations among transfer rates as a function of the strength of a measurement. These relations support the intuition that only quantum dynamics is susceptible to QZE. With the derived bounds on the magnitude of coherent dynamics, we propose an experimentally accessible coherence witness. Our results have potential application in assessing coherence of quantum transport in biological and other complex many-body systems.	
\end{abstract}

\maketitle

%\tableofcontents

\section{Introduction}

In the famous "Hitchhiker's guide to the Galaxy" \cite{adams_hitchhikers_1979} the author shows that formulating a proper question may turn out to be more difficult than finding its answer, be it 42 or anything else. In physical systems a well stated question on the character of a given system has a fundamental role.  Inspired by Douglas Adams we have approached the question of witnessing coherence in the evolution of an open system of which we have a limited knowledge and on which we are allowed to perform a limited set of measurements. The difficulty of formulating a proper question here comes from stating what is understood by quantum coherence, in general terms. To address the latter we have referred to an effect that is a signature of quantum behavior: quantum Zeno effect and from this we have derived our operational definition of coherence.

Characterization of open systems' dynamics is of general interest on its own. However, it becomes even more interesting when applied to systems that are known for their complexity. Prominent examples are many-body systems \cite{barontini_controlling_2013,barreiro_open-system_2011}, nanostructures~\cite{lambert_distinguishing_2010} and biological complexes~\cite{engel_evidence_2007,olaya-castro_efficiency_2008,ishizaki_theoretical_2009,chin_coherence_2012,chin_role_2013}. Recent debate on the presence of quantum coherence in certain biological complex systems~\cite{kassal_does_2013,gauger_sustained_2011,bandyopadhyay_quantum_2012} is a prime example that assessing \emph{coherence} and its character is far from straightforward.

These issues have been addressed from a variety of perspectives. In addition to the early studies of decoherence in quantum walks~\cite{mohseni_environment-assisted_2008}, a considerable amount of theoretical effort is devoted to understanding the role of quantum coherence in more general contexts, such as resource theories~\cite{cai_chemical_2013, baumgratz_quantifying_2013} and thermodynamics~\cite{rodriguez-rosario_thermodynamics_2013}. In addition, coherence witnesses of quantum states has been linked to the ability of describing the evolution of the system's populations in terms of stochastic propagators, namely, a necessary condition for the absence of quantum inferference~\cite{li_witnessing_2012}.

Regarding the assessment of coherence in physical evolution, several approaches have been put forward. Most notably, the methods of Quantum Process Tomography, --first developed with the profiling of hypothetical quantum computers in mind-- are now being exploited to assess coherence in biological complexes~\cite{yuen-zhou_coherent_2013}. Other proposals include the Leggett-Garg inequality~\cite{leggett_quantum_1985,wilde_could_2010,chen_examining_2013}, temporal CHSH inequalities~\cite{brukner_quantum_2004,fritz_quantum_2010} and the so-called \emph{no-signalling in time} condition~\cite{kofler_condition_2013}. These proposals rely in one way or another in assuming that classical systems can be subject to measurement without perturbation.

In this article, we draw a quantitative link between quantum coherence and the Quantum Zeno effect~\cite{misra_zenos_1977, facchi_quantum_2002, facchi_quantum_2008}. Our approach draws from the intuition that the Quantum Zeno effect has a close relationship with the quadratic buildup of probabilities of unitary evolution. In contrast to quantum evolution, it is well understood that classical rate equations are not subject to the QZE even in the presence of measurement back-action. Hence, one may expect that the extent to which a system's dynamics can be subjected to Quantum Zeno effect is an indicator of the amount of coherence present in the evolution. We provide a rigorous quantitative formulation of these ideas. In addition, we show that quantum coherence of physical evolution can be assessed with a minimal set of state preparations and a single measurement setup, providing coherence witnesses potentially tight. Our approach should be regarded as a proof-of-principle, showcasing the essential features of these ideas, but open to a wide variety of extensions tailored to the particular systems or contexts of interest.

The paper is organized as follows. In Section~\ref{framework} we define the problem and outline the main result of our work, stating without proof several facts that will be discussed in greater detail in the following sections. In Section~\ref{measurement} we describe in detail our measurement protocol, and discuss the experimentally accessible quantities that will be used to extract our coherence witness. Section~\ref{master} derives the effective dynamics within our approximations, by performing an exact adiabatic elimination of population transfer via coherent transitions. This provides a generic closed form for the measured quantities, that will be used in Section~\ref{bounds} to extract the witnesses that we are interested in. Section~\ref{sec:examples} illustrates our procedure with three examples, and discusses the performance of our bounds. Section~\ref{sec:discussion} concludes with some general observations and suggests future lines of work.

\section{General framework and main results}\label{framework}

Our analysis is concerned with a generic finite-dimensional quantum system described by a finite-dimensional Hilbert space $\H$. We denote subspaces of $\H$ by script uppercase symbols, {\em e.g.} $\PP$. The most general time-independent evolution in the Markovian approximation is described by the master equation 
\begin{subequations}
\label{meq}
\begin{align}
	\frac{d}{d t}\rho(t)=\mathcal L[\rho(t)],
\end{align}
where $\L$ is a time-independent, Lindblad-Kossakowski superoperator~\cite{kossakowski_quantum_1972, lindblad_generators_1976}
\begin{align}\label{eq:lindblad}
	\mathcal L[\rho]=-i[H,\rho]+\sum_\mu\Big(W_\mu\rho W_\mu^\dagger-\frac{1}{2}\{W^\dagger_\mu W_\mu,\rho\} \Big).
\end{align}
\end{subequations}

Here $H$ stands for a Hamiltonian and $\{W_\mu\}$ is a set of jump operators representing the noise.
As a general rule, we will write linear operators on $\H$ with Greek or upper case symbols, and  superoperators (linear maps thereof) will be written in calligraphic letters ({\em e.g.} $\P$) with the single exception of $\phi$, which denotes a completely-positive superoperator.

Given $\L$ and a set of jump operators $\{W_\mu\}$, Hamiltonian $H$ is uniquely determined. Here we do not assume detailed knowledge of jump operators, but we assume certain property of noisy mechanism at work and focus on the question as to how to obtain information on the {\em missing} $H$. Cases when we cannot uniquely determine the Hamiltonian part of $\L$ will be mentioned later.

%{\color{red} It is important to note that a given $\L$ superoperator is not uniquely represented by expressions of the form \eqref{eq:lindblad} or \eqref{eq:lindblad2}~\cite{lindblad_quantum_1991, tarasov_quantum_2008}. This will be discussed in Section~\ref{decoherence}. Notice, however, that }

We focus our attention on systems for which there exists a set of $n$ orthogonal projections $\{P_i\}$, $\sum_{i=1}^n P_i=\openone$, $P_iP_j=P_i\delta_{ij}$, such that
\begin{align}\label{eq:phiconditions}
	\sum_{ r\neq s}\sum_\mu P_r W_\mu P_iW_\mu^\dagger P_s=\sum_{ r\neq s}\sum_\mu P_r W_\mu^\dagger P_iW_\mu P_s=0,
\end{align}
for all $i$. Intuitively, this condition enforces that there exists a decomposition of the system's Hilbert space  
\begin{align}\label{decomp}
	\H=\bigoplus_{i=1}^n \PP_i,
\end{align}
with respect to which the incoherent mechanism does not create coherence, and neither does its adjoint --incoherent states remain incoherent in the Schroedinger picture, and so do observables in the Heisenberg picture--. This is the only property of the noisy evolution that we assume to be known. In Section \ref{sec:examples} we will provide natural examples where this is the case. 

Our work aims at providing experimentally accessible measures of coherence obtained by the outcomes of a unique measurement characterized by projectors $\{P_i\}$. By implementing a Hamiltonian $k H_m=k \sum_i \eta_i P_i$ in addition to the system's dynamics $\L$, the coherent coupling among subspaces $\PP_i$ is suppressed at leading order in $1/k$. The actual transition probabilities are mediated, at leading order $(\O(1))$, by the effective incoherent dynamics derived from the dissipative part of~$\L$. At subleading order $\O(1/k)$ the population transfer among $H_m$'s eigenstates is mediated via virtual transitions through superposition states. It will be this dependence on $k$ that will reveal the contribution of quantum coherence to the observed population transfer rates.

We consider two types of coherence measures: on the level of decomposition~\eqref{decomp}, we consider the off-diagonal blocks $H_{ij}=P_iHP_j$ and their 2-norm $\|H_{ij}\|_2=(\tr[H_{ij}^\dg\, H_{ij}])^{1/2}$ as a measure of how strongly subspaces $\PP_i$ and $\PP_j$ are coupled. These quantities are readily accessible from our formalism. In addition, our scheme provides lower bounds to the spectral spread of the system's Hamiltonian $H$, 
\begin{align}
	\coh(H)\equiv \lambda_{\max_{}}(H)-\lambda_{\min_{}}(H),
\end{align}
where $\{\lambda_i(H)\}$ represent $H$'s eigenvalues. 

Consider the probabilities of preparing the system in states $\rho_j=P_j/\dim\PP_j$ at time $0$ and obtianing outcome $P_i$ at time $t$, when the coupling strength is set at $k$. These probabilities are arranged in $[\sfP(k,t)]_{ij}$. As a general rule all matrices of observable magnitudes derived from $\sfP$ will be written in sans-serif caption. With a suitably chosen tensor $\sfW_{ijk}$ whose components depend on the chosen values of $k$, the time $t$ and the eigenvalues $\eta_i$ of $H_m$, we have that for $i\neq j$
\begin{align}\label{eq:result1}
	\|H_{ij}\|_2=\sqrt{\dim\PP_j\sum_{k\in K}\sfW_{ijk}[\dot\sfP(k,t)]_{ij}},
\end{align}
with $\dot\sfP$ denoting the probabilities' derivatives w.r.t. time, and the sum over $k$ running over a finite set of values $K$. This provides a quantifier for the size of each off-diagonal block in the Hamiltonian. In the ideal case of decomposition~\eqref{decomp} into one-dimensional subspaces, this procedure determines the norm of each off-diagonal entry in $H$. This information is sufficient, in itself to bound $\coh(H)$ away from zero. However, the $\|H_{ij}\|$ do not contain all the information present in $\dot\sfP$. In the following we provide a lower bound based on the latter.

A coherence witness for $\coh(H)$ is an experimentally accessible quantity $\Omega$ such that
\begin{align}
	\Omega\leq \coh(H).
\end{align}
As we will show, the following is a coherence witness for $\coh(H)$,
\begin{align}\label{eq:result2}
	\Omega=\bigg\|\Big(\sum_{i\neq j}\sum_{k\in K}\sfW_{ijk}\sfD^{-1/2}\dot\sfP(k,t)\sfD^{1/2}\Big)^{1/2}\bigg\|_\infty,
\end{align}
where $\|X\|_\infty$ denotes the operator bound norm, {\em i.e.} $X$'s largest eigenvalue, $\sfD=\diag(\dim \PP_1,\ldots,\dim \PP_n)$ and $K$ is a finite set of values of parameter $k$.

In deriving this inequality, we will show that it is the system's susceptibility to Zeno effect, namely, the potential for altering the system's dynamics by continuous measurement, that reveals the presence of a coherent contribution to the system's evolution.

The present work is a proof-of-principle for establishing noise-independent lower bounds on $\coh(H)$ and estimates of $\|H_{ij}\|_2$ with minimal assumptions and preparation/measurement setups. 

\section{Measurement scheme}\label{measurement}

We begin with a generic master equation of the form Eq.~\eqref{meq} with time-independent generator $\L$, subject to an additional controllable Hamiltonian 
\begin{align}\label{eq:measHam}
	H_m=\sum_{i=1}^n \eta_i P_i,
\end{align}
which we call {\em measurement} Hamiltonian. We will use an overall parameter $k$ to denote the intensity of this, and thus regard $H_m$ as dimensionless.

We use Hamiltonian $H_m$ to induce a continuous coherent driving, with $H_m$ having no more degeneracy than that imposed by the ranks of the projectors $P_i$, ($\eta_i=\eta_j\Leftrightarrow i=j$). The effective result of this driving is an induced quantum Zeno effect on the coherent part of the system's dynamics. With our setup, the Zeno subspaces are $\{\PP_i\}$, and we make no assumptions on their dimensions other than being known, $d_i\,\equiv\,\tr P_i$. It is not necessary for us to assume any specific values of eigenvalues $\eta_i$, however, from the perspective of the technique used in the course of this work, our preferred choice of the measurement design is to make the differences $(\eta_i\,-\,\eta_j)$ unique.  

The dynamics of the system is given by the equation
\begin{align}\label{eq:mastereq}
	\frac{d}{d t}\rho(t)=\mathcal L[\rho(t)]\,-\,ik[H_m,\rho(t)].
\end{align}
%At large coupling strength $k$ the effective dynamics approaches the Zeno dynamics, and the corrections contain the adiabatically eliminated transitions through coherent coupling. 
We assume that the frequencies related to $kH_m$ are not faster nor comparable to the frequencies related to the processes underlying decoherence in the dynamics given by jump operators $\{W_\mu\}$. In other words, the magnitude of $kH_m$ does not conflict with the Markovian approximation underlying Eqs.~\eqref{meq}. In the following, when we refer to the large $k$ limit, and denote it by $k\rightarrow\infty$, one must bear in mind that this limit is constrained within the validity of the Markovian approximation underlying Eq.~\eqref{meq}. With this consideration in mind, we can safely assume that the Lindblad representation of the evolution is valid throughout the entire measurement and the system's dynamics remains Markovian~\cite{gardiner_quantum_2004}.

We consider the system initialized in the maximally mixed state in one of the measurement subspaces, namely, $\rho(0)\,=\,\rho_i\,\equiv\,P_i/d_i$. Next, we introduce the projection superoperator, a centralizer of $H_m$, defined as 
\begin{equation}
\P[\rho]\,\equiv\,\sum_iP_i\rho P_i,
\end{equation}
along with its complementary projector
\begin{equation}
\Q[\rho]\,\equiv\,(\id\,-\,\P)[\rho]\,=\,\sum_{k\neq l}P_k\rho P_l,
\end{equation}
where $\id$ is the identity superoperator. With the above mentioned choice of the initial state, we satisfy $\rho(0)\,=\,\P[\rho(0)]$. 

We choose the measurement protocol to be the following: 

\begin{enumerate}
\item {\bf Preparation: } The system is prepared at time $0$ in one of the measurement subspaces: $\rho(0)\,=\,\rho_j$,
\item {\bf Evolution:} Let the system evolve for the appropriately chosen small time $t$, with continuous driving $kH_m$ (with strength $k$). 
\item {\bf Measurement: } At time $t$ a conclusive projective measurement $\{P_i\}$ is performed.
\item {\bf Estimation:}  Repetition of this process with different initial preparations yields the probabilities of finding outcome $i$ at time $t$, when the system was prepared at time $0$ in state $j$, and the evolution is continuously driven with $H_m$ at strength $k$.
\end{enumerate}
This procedure yields the generalized transition probabilities $p_{i\leftarrow j}(k,t)$, which can be conveniently arranged in a matrix $[\sfP(k,t)]_{ij}=p_{i\leftarrow j}(k,t)$. By measuring  $\sfP(k,t)$ at various times and coupling strength values, $k$, one obtains sufficient information about the dynamics to be able to place lower bounds to the amount of coherence and decoherence present in the dynamics. We regard the rates $[\mathsf{\dot P}(k,t)]_{ij}=\frac{d}{dt}p_{i\leftarrow j}(k,t)$ as the time-derivative of the transition probabilities
\begin{equation}\label{eq:Rdef}
	\mathsf{\dot P}_{ij}^{(k)}\,\equiv\,[\mathsf{\dot P}(k,t)]_{ij}\,\equiv\,\frac{d}{dt}p_{ij}(k,t).
\end{equation}
Our set of experimental data will consist of transition rates between various measurement subspaces $(i,j)$ for a set of measurement strengths $K=\{k_1,\ldots,k_N\}$, measured at appropriately chosen small time $t$. 

The next requirement in our analysis is to establish an analytical correspondence between transition rates and specific properties of the Lindblad superoperator that we are interested in.

\section{Effective dynamics}\label{master}

Now we obtain the dynamics of the centralized density operator $\P[\rho(t)]$ at suitably chosen small times $t$, in terms of the initial state $\rho(0)\,=\,\P[\rho]$ and the driving strength $k$.
It is convenient for our purposes to write $\L$ as a combination of two terms,
\begin{align}\label{eq:lindblad2}
	\frac{d}{dt}\rho=-i\ad_H[\rho]+\L_\phi[\rho],
\end{align}
where we have defined
\begin{subequations}
\begin{align}
	\ad_H[\rho]=&\,[H,\rho],\\
\label{dissip}
	\L_\phi[\rho]=&\,\phi[\rho]-\frac{1}{2}\{\phi^*(\openone),\rho\}.
\end{align}
\end{subequations}
Here, $\ad_H$ is the {\em adjoint action} well known in the theory of Lie algebras~\cite{fulton_representation_1991}, and $\phi$ is a completely-positive map,
\begin{align}
	\phi(\rho)=&\sum_\mu W_\mu \rho W_\mu^\dagger.
\end{align}
In \eqref{dissip}  $\phi^*=\sum_\mu W_\mu^\dagger\,\cdot\,W_\mu$ denotes the adjoint of $\phi$ with respect to the Hilbert-Schmidt inner product, {\em i.e.}, $\tr[X^\dg \phi(Y)]=\tr[(\phi^*(X))^\dg Y]$.

We start with the generalized Liouville equation~\cite{breuer_theory_2002} including $kH_m$, where $k$ controls the strength of the driving mechanism,
\begin{equation}\label{eq:motherME}
\frac{d}{dt}\rho(t)\,=\,\L[\rho(t)]\,-\,ik \,\ad_{H_m}[\rho(t)].
\end{equation}
We are interested in the dynamics of the system for small times $t\ll 1/\| \L\|$ as compared with the typical timescales of the Lindblad generator. %In particular, we are interested in the evolution of the population operator $\P[\rho(t)]$, which contains the probabilities of obtaining outcome $\eta_i$ upon measurement of $H_m$. 
The details of the derivation are contained in Appendix~\ref{app:TCL}. At next-to-leading order in $t$ we have
%
%
%We move to the interaction picture with respect to $H_m$, define $U(t)\,=\,\exp i t\ad_{H_m}$ and write down equations for both $\frac{d}{dt}\P[\rho(t)]$ and $\frac{d}{dt}\Q[\rho(t)]$. This is achieved by applying projector operator techniques to the Liouville equation, $\frac{d}{dt}\P[\rho(t)]$ being the part of the density operator that we are interested in. Note that contrary to the standard master equation derivation, we do not use the usual projection onto the system's and bath's product state; instead, we project the density operator onto the measurement subspaces. 
%
%Next, we look at the dynamics at small times and define a small parameter of the perturbative expansion, $\epsilon\,\equiv\,t\|\L\|\,\ll\,1$. Following first steps of the time-convolutionless master equation approach~\cite{breuer_theory_2002} we can write (back in the Schroedinger picture)
%
%\begin{equation}
%\frac{d}{dt}\P[\rho^I(t)]\,=\,\big(\P\L^I\P\,+\,\P\L^I\Q\big)[\rho^I(t)],
%\end{equation}
\begin{align}
&\frac{d}{dt}\P[\rho(t)]=\\
\nonumber&=\P\L\left(\P+\int_0^tds \,\U(s-t)\L\P+\mathcal{O}\left(t^2\|\L\|^2\right)\right)[\rho(0)],
\end{align}
where $\U(t)[x]=e^{itk\,H_m}xe^{-itk\,H_m}$.
One can readily see that first term $\P\L\P=-i\P\ad_{H_Z}\P+\P\L_\phi\P$ gives rise to the effective Zeno Hamiltonian $H_z=\P[H]$, together with the effective dissipative dynamics among subspaces $\{\PP_i\}$. The second term contains the adiabatically eliminated population transfer due to coherences, which occur only at next-to-leading order in the strong driving $kH_m$. This expression is amenable to exact integration, thus yielding
\begin{subequations}\label{me}
\begin{align}
\frac{d}{dt}\P[\rho(t)]=\big(\D_0(t)+\D_1(t)+\L\times\mathcal{O}(t^2\|\L\|^2)\big)[\rho(0)],\label{me1}
\end{align}
with
\begin{eqnarray}
	\D_0(t)&\equiv&\P\L\P(\mathcal I+t\P\L\P),\\
\label{D1}
	\D_1(t)&\equiv&\P\L\Q\frac{\mathcal I-e^{-i tk\,\ad_{H_m}}}{ik\ad_{H_m}}\Q\L\P.
\end{eqnarray}
\end{subequations}%
This equation is valid for small times defined by $t\ll 1/\|\L\|$ and for all values of $k$. We neglect the term $\P\L\times\mathcal{O}(t^2\|\L\|^2)$ as it is subleading w.r.t. the other terms. The detailed derivation of Eq.~\eqref{me} can be found in Appendix~\ref{app:TCL}.

Note that when we take $k\rightarrow\infty$, the master equation reduces to
\begin{equation}\label{meqZeno}
\frac{d}{dt}\P[\rho(t)]\,=\,\D_0(t)[\rho(0)]\,=\,\P\L\P(\mathcal I\,+\,t\P\L\P)[\rho(0)].
\end{equation}
The above can be understood as an effective (Zeno) dynamics characterized by $\L_Z=\P\L\P$, which acting on centralized states $\rho(0)=\P[\rho(0)]$ can be expressed as%
\begin{eqnarray}
\L_Z=-i\ad_{H_Z}\,+\L_{\phi_\eff}\label{effL}
%\,-\,\frac{1}{2}\{\phi_\eff^*(\openone),\cdot\},
\end{eqnarray}
where $H_Z=\P[H]$ is the Zeno Hamiltonian, $\phi_\eff=\P\phi\P$ describes the effective decoherence process and the order $t$ term in Eq.~\eqref{meqZeno} corresponds to the first term in the expansion $\rho(t)=\exp(t\L_Z)\rho(0)$.

Our main interest lies in the operator $\D_1(t)$ which \emph{a)} depends on $k\,H_m$, \emph{b)} couples subspaces defined by $\P$ and $\Q$ and \emph{c)} imprints phases onto the $(i,j)$ blocks in $\Q$. Notice that the adjoint action $\ad_{H_m}$ has support on the subspace defined by $\Q$, thus the expression $\Q\ad_{H_m}^{-1}\Q$ is well defined, and can be written as
\begin{align}
	\Q\ad_{H_m}^{-1}\Q[X]=\sum_{i\neq j} \frac{1}{\eta_i-\eta_j}P_i X P_j,
\end{align}
with immediate generalization to the expression encountered in Eq.~\eqref{D1}.

Superoperators $\D_0$ and $\D_1$ capture the essential physics revealed by the measurements, and the population transfer rates (Eq.~\eqref{eq:Rdef}) are given by
\begin{align}
	[\dot \sfP(k,t)]_{ij}=\tr\Big[P_i\,\big(\D_0(t)\,+\,\D_1(t)\big)[\rho_j]\Big].
\end{align}

To discuss the consequences of both $\D_0(t)$ and $\D_1(t)$ we need to treat them on equal footing. Therefore we need to guarantee that the second order term in $\D_0(t)$ can be compared with $\D_1(t)$, due to the magnitudes of $t\L$ and $k$. We find that $kt\sim1$ is a suitable regime of $k$ to work with.

\section{Bounds on coherence}\label{bounds}

We are now in a position where we can derive bounds for coherence measures of $\ad_H$, and thus, come to the main results of the present work. As has been discussed in the introduction, in the strong driving regime (Zeno regime, $k\rightarrow \infty$) all coherent population transfer between the Zeno subspaces is suppressed, and the remaining dynamics between those can ultimately be attributed to the incoherent processes of the system. We illustrate this with examples in Section~\ref{sec:examples}. This does not mean that incoherent dynamics is unaffected by the measurement. As shown in Eq.~\eqref{effL} the map $\phi$ describing the incoherent process is also modified, but it remains relevant as long as population transfer is regarded.

%A simple example is the case in which no stationary state of $\L$ commutes with the decoupling Hamiltonian $H_m$. In such case, the decoherence mechanism is unable to drive the system to a stationary state and an effective decoherence emerges as illustrated by Eq.~\eqref{meqZeno}. {\color{green} How this example is related to our assumption about the jump operators?}

%This effect, although it can never be as dramatic as in the coherent case --total suppression of dynamics--, implies that the observed rates can only provide partial information about the strength of the incoherent dynamics. 

%Nevertheless, the resulting effective dissipation dynamics in the Zeno regime, will have some general features which can be exploited to extract the subleading behavior, namely, that dictated by $\D_1$, of which virtual coherent processes are responsible for the population transfer.

The main observation which will be recurring in the following is that the rate matrix, $\dot \sfP$, can be regarded as a minor of the matrix representation of the superoperator $\D_0+\D_1$ in a suitably chosen basis of $L(\mathcal H)$, $\{P_1,\ldots,P_n,T_1,\ldots T_{d^2-n}\}$, where $T$'s complete the basis defined by $\{P_i\}$. Since this basis is not orthonormal, we introduce the orthonormal operators $\tilde P_i=P_i/\sqrt{d_i}$, so that
\begin{align}
	\tr[\tilde P_i \tilde P_j]=\frac{\tr[P_i P_j]}{(d_id_j)^{1/2}}=\delta_{ij}.
\end{align}
In addition, we arrange dimensions $d_i$ in the matrix $\sfD=\diag(d_1,\ldots,d_n)$, so that we can define the {\em normalized} rate matrix $\sfR$
\begin{align}\label{eq:R}
	\sfR^{(k)}(t) = \sfD^{-1/2} \dot\sfP(k,t)\sfD^{1/2}.
\end{align}
The normalized rates $\sfR_{ij}^{(k)}(t)$ can be written as %
\begin{equation}
	\sfR_{ij}^{(k)}(t)=\tr\big[\tilde P_i\big(\D_0(t)+\D_1(t)\big)[\tilde P_j]\big].
\end{equation} 
Recalling that $\Q(\cdot)\,=\,\sum_{r\neq s}P_r(\cdot)P_s$ and $H_mP_r\,=\,\eta_rP_r$ we can write 
\begin{equation}
	\sfR_{ij}^{(k)}=\tr\bigg[\tilde P_i\,\Big(\D_0(t)\,+\,\sum_{r\neq s} \frac{1-e^{i kt (\eta_r-\eta_s)}}{i k (\eta_r-\eta_s)}\P\L\Q_{rs}\L\P\Big)[\tilde P_j]\bigg],
\end{equation}
where $\Q_{rs}[\cdot]=P_r\,\cdot\,P_s$ projects on a specific off-diagonal block corresponding to the pair $(r,s)$,
\begin{align}
	\Q_{rs}[\rho]=\left(\begin{array}{cccccc}
		0&\cdots& 0&\cdots&0\\
		\vdots&&\vdots&&\vdots\\
		0&\cdots&P_r\rho P_s&\cdots&0\\
		\vdots&&\vdots&&\vdots\\
		0&\cdots&0&\cdots&0\\
	\end{array}\right).
\end{align}
With the right choice of $\{\eta_r\}$, the differences $\eta_r-\eta_s$ are unique, and the off-diagonal blocks $\Q_{rs}$ gain unique frequencies. Next, we recover the matrix representation of $\P\L\Q_{rs}\L\P$ by a suitable linear transform.

%As discussed in Section~\ref{master}, the coherent character of $\L$ is imprinted on the subleading contribution, $\D_1(t)$, and thus it is the main interest of the present work to show how this can be bounded away from zero. 

The contribution to the $\sfR$ matrix away from the Zeno regime ($k\rightarrow\infty$) has been shown to be $\sfR^{(k)}=\sfR^{(Z)}+\sfR^{(1)}$, where $\sfR^{(Z)}$, $\sfR^{(1)}$ are given by
\begin{subequations}
\begin{eqnarray}\label{eq:R1}
	\sfR^{(Z)}_{ij}&=&\tr[\tilde P_i\L(\id-t\P\L)[\tilde P_j]]\\
	\sfR^{(1)}_{ij}(k)&=&-i\, \tr[\tilde P_i\, \P\L\Q\frac{\id-e^{-itk\,\ad_{H_m}}}{k\,\ad_{H_m}}\Q\L\P[\tilde P_j]].~~~~~~~
\end{eqnarray}
\end{subequations}
In order to recover the details of the superoperator $\ad_H$, it is convenient to work in the superoperator eigenbasis of $\ad_{H_m}$. Operators of the form $P_r O P_s= O$ span the eigenspace of $\ad_{H_m}$ with eigenvalue $\omega_{rs}=\eta_r-\eta_s$. Namely, eigenvectors consist of operators with nonzero entries liying in a unique off-diagonal block, labeled by row-column indices $(r,s)$. We label pairs of indices with Greek letters, so that if $(r,s)=\mu$ we write $\omega_\mu$ for the frequencies in $\D_1$ and $Q_\mu[\rho]=P_r\rho P_s$. Notice that $\ker(\ad_{H_m})=\ker\Q$, so that the expression $\Q(1-\U^*(t))\ad_{H_m}^{-1}\Q$ appearing in Eq.~\eqref{eq:R1} is well defined even though $\ad_{H_m}$ is singular. Defining 
\begin{align}
	\mathsf{M}_{k\mu}= \frac{e^{-itk\omega_\mu}-1}{ik\omega_\mu}
\end{align}
we can write $\sfR^{(1)}$ as
\begin{align}
	\sfR^{(1)}_{ij}(k)=-\sum_\mu \mathsf{M}_{k\mu}\tr[\tilde P_i\,\P\L\Q_\mu \L\P[\tilde P_j]].
\end{align}
The matrix $\mathsf{M}$ is indexed by the different values of $k\in K$ and the frequencies $\mu=\{1,\ldots,n(n-1)\}$. Therefore $\mathsf{M}$ is not necessarily a square matrix, and the number of rows depends on our choice of how many values of $k$ are sampled. Let us assume that we chose to sample one more value of $k$ than that of frequencies, {\em i.e.}, $|K|=n(n-1)+1$, thus $k\in K=\{k_0,k_1,\ldots,k_{n(n-1)}\}$.

Let us introduce the pseudoinverse of $\mathsf{M}$, $\mathsf{W}$ such that
\begin{subequations}\label{inversetransform}
\begin{align}
\label{it1}	\sum_{k\in K} \mathsf{W}_{\mu k}\mathsf{M}_{k\nu}\,&=\delta_{\mu\nu},\\
\label{it2}	\sum_{k\in K} \mathsf{W}_{\mu k}\,&=0.
\end{align}
\end{subequations}
This can be always satisfied when $|K|=n(n-1)+1$ by taking the singular value decomposition of $\mathsf{M}$,
\begin{align}
	\mathsf{M}=U \left(\begin{array}{c}0\\ \hline S\end{array}\right)V
\end{align}
where we have highlighted the lower $n(n-1)\times n(n-1)$ block corresponding to the $S=\diag(s_1,\ldots s_{n(n-1)})$ singular values of $\mathsf{M}$. Let $u_{kk'}$ be the matrix elements of $U$. Then the choice 
\begin{align}
	\mathsf{W}=V^\dagger\left(x\Big|S^{-1}\right)U^\dagger
\end{align}
where $x$ is a vector defined as 
\begin{align}\label{x}
	x_\mu=-\frac{\sum_k u^*_{k\mu}}{s_\mu \sum_k u^*_{k0}},\quad\mu=1,\ldots,n(n-1),
\end{align}
is a solution to Eqs.~\eqref{inversetransform}. A few words of caution are in order. In choosing the energy levels of $H_m$ and values of $k$, one must ensure that $\mathsf{M}$ has $n(n-1)$ nonzero singular values. Otherwise Eq.~\eqref{x} is ill-defined. Fortunately, the freedom in choosing these values is large enough to make sure this case is always avoidable.

Having at hand the pseudoinverse $\mathsf{W}$ one can use it to obtain
\begin{eqnarray}\label{eq:T}
	\sfT_\mu&=&\sum_k \mathsf{W}_{\mu k}\sfR^{(k)}\\
	&=&\sum_k \mathsf{W}_{\mu k}[\sfR^{(Z)}+\sfR^{(1)}(k)]
\end{eqnarray}
The first sum vanishes due to Eq.~\eqref{it2} leading to
\begin{align}
	[\sfT_\mu]_{ij}=-\tr[\tilde P_i\,\P\L\Q_\mu \L\P[\tilde P_j]].
\end{align}
The $\sfT_\mu$ matrix may be called {\em Zeno susceptibility} as it determines how the system's frequency $\omega_\mu$ responds under the Zeno measurement, or continuous driving. 
Finally, by taking into account assumptions discussed in the beginning of this work \eqref{eq:phiconditions}, one can show that [See Appendix~\ref{app:phidrops}]
\begin{subequations}\label{eq:PLQzero}
\begin{align}
	\Q\L_\phi\P[P_i]&=0\\
	\Q\L_\phi^*\P[P_i]&=0,
\end{align}
\end{subequations}
so that $\sfT_\mu$ can be written as
\begin{align}\label{eq:Cmu}
	[\sfT_\mu]_{ij}=\tr[\tilde P_i\,\ad_H\Q_\mu \ad_H[\tilde P_j]].
\end{align}
$\sfT_\mu$ is the matrix representation of $\ad_H\Q_\mu\ad_H$ within the subspace of $L(\mathcal H)$ spanned by $\{P_i\}$. These matrices contain the essential information which we are interested in. The motivation of assumption \eqref{eq:phiconditions} --or equivalently \eqref{eq:PLQzero}-- is now clear. It ensures that incoherent dynamics does not couple Zeno subspaces  in the adiabatically eliminated virtual --second order-- transitions.

\subsection{Coherent coupling between Zeno subspaces}
%The induced norm of $\ad_H$ 

The magnitude of $\coh(H)$ characterizes the fastest timescales at which the system can coherently evolve and will concern us later. However, as will be seen in Section~\ref{sec:examples}, the measurement specified by $\{P_i\}$ often represents a physically meaningful decomposition of the system's Hilbert space. It is therefore of interest to quantify the coupling  among subspaces $\PP_i$ induced by the system's Hamiltonian $H$. Consider a pair of subspaces $\PP_i$, $\PP_j$, and the corresponding block in the Hamiltonian, given by the operator $H_{ij}=P_i HP_j$. The norm on the latter immediately quantifies the strength of the coupling between $\PP_i$ and $\PP_j$. In particular, the Hilbert-Schmidt norm, $\|X\|_2=\sqrt{\tr[XX^\dagger]}$, is directly related to Zeno susceptibility $\sfT_\mu$. Indeed, an easy calculation shows that, for $i\neq j$
\begin{align}\label{eq:h}
	\sqrt{(d_id_j)^{1/2}[\sfT_{ij}]_{ij}}=\|H_{ij}\|_2
%	\|h_{ij}\|_2= \sqrt{(d_id_j)^{1/2}}
\end{align}
which shows that the 2-norm of $H_{ij}$ is readily available from our measurement scheme. Eq.~\eqref{eq:h} combined with Eqs.~\eqref{eq:T} and \eqref{eq:R} gives the claimed result, Eq.~\eqref{eq:result1}.

The singular value decomposition of $H_{ij}$ suggests that there are bases in $\PP_i$, $\PP_j$  such that [supposing $\dim(\P_i)>\dim(\P_j)$]
\begin{align}
	H_{ij}=U\left(\begin{array}{cccccc}
		s_1\\
		&s_2\\
		&&\ddots\\
		\hline\\
		&0\\
		&
	\end{array}\right)V^\dagger
\end{align}
with $U$, $V$ unitaries in $\PP_i$, $\PP_j$ respectively. Hence the singular values of $h_{ij}$ are the largest coupling strengths that two orthogonal sets of vectors in $\PP_i$, $\PP_j$ can have. Thus, $\|H_{ij}\|_2=\sqrt{\sum_n s_n^2}$ measures the coherent coupling strength between subspaces $\PP_i$ and~$\PP_j$. 
%%%%%%%%%%%%%%%%%%%
%%%%%%%%%%%%%%%%%%%
%%%%%%%%%%%%%%%%%%%
\subsection{A universal measure of coherence}\label{sec:universal_coh}
%%%%%%%%%%%%%%%%%%%
%%%%%%%%%%%%%%%%%%%
%%%%%%%%%%%%%%%%%%%
The magnitudes $\|H_{ij}\|_2$ characterize the coupling strengths among Zeno subspaces $\PP_i$, $\PP_j$, but one may be interested in obtaining estimates of the overall total strength of the Hamiltonian $H$. This is of course not always possible due to couplings occurring within any given Zeno subspace, which are not accessible to our measurement scheme. However, the dependence of the rates $\sfR$ on $k$ can provide nontrivial lower bounds to $\coh(H)$.

Summing Eq.~\eqref{eq:Cmu} over all distinct pairs $\mu=(r,s)$, $r\neq s$ we get
\begin{align}\label{eq:C}
	\sfC_{ij}=\sum_\mu [\sfT_\mu]_{ij}=\tr\Big[\tilde P_i \ad_H \Q\ad_H [\tilde P_j]\Big].
\end{align}
Observing that $\Q\ad_H[P_j]=\ad_H[P_j]$, --namely, commutators only couple populations to coherences and viceversa, so the presence of $\tilde P_j$ allows to remove the coherence projector $\Q$,-- leads to the conclusion
\begin{align}
	\sfC_{ij}=\tr\big[\ad_H[\tilde P_i]^\dagger \,\ad_H[\tilde P_j]\big].
\end{align}
Clearly $\sfC$ is positive semidefinite, thus it has a well defined square root. Let us define its operator norm
\begin{align}
	\Omega=\|\sqrt\sfC\|_\infty.
\end{align}
Exploiting the monotonicity of the operator norm [See Appendix~\ref{app:monotonicity}] we show that $\Omega$ is a lower bound to the induced Hilbert-Schmidt superoperator norm of $\ad_H$
\begin{align}
	\Omega \leq \max_{\|X\|_2=1}\big\| \ad_H[X]\big\|_2\equiv\|\ad_H\|_2.
\end{align}
The induced Hilbert-Schmidt norm is $\|\ad_H\|_2=\coh(H)$ [see Appendix~\ref{app:opnorm}],\begin{align}\label{eq:mainresult}
	\Omega \leq|E_{\max_{}}-E_{\min_{}}|=\coh(H),
\end{align}
where $E_{\max_{}}$, $E_{\min_{}}$ are the highest and lowest energy eigenvalues respectively. This combined with Eqs.~\eqref{eq:C}, \eqref{eq:T} and \eqref{eq:R} yields our main result, Eq.~\eqref{eq:result2}. A nonzero value of $\Omega$ is a witness that there is a nontrivial Hamiltonian contributing to the dynamics. This, in turn is an indicator that the dynamics of the system cannot be explained solely in terms of classical rate equations.

More precisely, the experimentally accessible $\Omega$ provides a lower bound on the spectral spread of $H$, $\coh(H)$. Notice that in obtaining this bound, only generic assumptions of the dissipation are made, and no requirement is put on its strength. Naturally, if $\|\L\|$ is very large, the timescales $t$ at which the system needs to be measured become small; this may be due to very high decoherence rates. However, given the order of magnitude of the $\|\L\|$, and a properly chosen small time $t$, the resulting bounds are independent of the details of the decoherence process or its actual strength.

\section{Examples}\label{sec:examples}

\subsection{A qubit undergoing Rabi oscillations}
As a first example we consider a simple two level system undergoing a spontaneous emission type incoherent process. We use the Pauli basis to write
\begin{align}
H&=\frac{\Delta}{2}(\cos\theta\,\sigma_z+\sin\theta\,\sigma_x),\\
\L[\rho]&=-i[H,\rho]+\gamma\big(\sigma^-\rho\sigma^+-\frac{1}{2}\{\sigma^+\sigma^-,\rho\}\big),\\
H_m&=\eta_1\ket{1}\bra{1}\,+\,\eta_2\ket{2}\bra{2},
\end{align}
where $\sigma^-\,=\,|0\rangle\langle 1|$ and $E,\theta$ and $\gamma$ are parameters of the model. Our measures of coherence in this model are easily shown to be
\begin{align}
	\|H_{12}\|_2&=\frac{1}{2}|\Delta\sin\theta|,\\
	\coh(H)&=|\Delta|.
\end{align}
In addition, there is only one nontrivial decomposition~Eq.~\eqref{decomp}, consistent with Eq.~\eqref{eq:phiconditions}, namely: $P_1=\ket1\bra1$ and $P_2=\ket2\bra2$.

\begin{figure}
%\begin{psfrags} % Optional
%	\psfrag{th}[b][b]{$\theta$}
%	\psfrag{C}[b][b]{$\coh(H)$}
	\includegraphics[width=\linewidth]{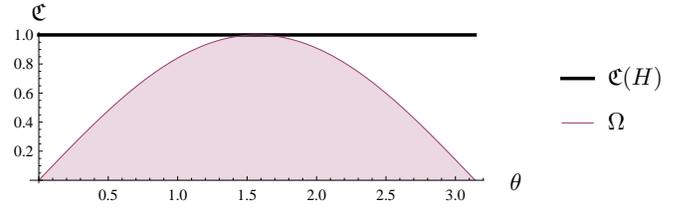}
%\end{psfrags} % Optional
\caption{\label{fig:qubit}Plot for the coherence measure $\coh$ (upper) and its lower bound, as a function of parameter $\theta$. The outer line represents the constant value of $\coh=|\Delta|$, which has been normalized to $1$. The inner line represents the bound obtained from the measurement protocol. Clearly, when $H$ is proportional to $\sigma_z$ ($\theta=0,\pi$), no possible Zeno effect can be induced and the resulting lower bound is zero. The bound is tight when the Zeno dynamics completely suppress the coherent population transfer ($\theta=\pi/2$).}
\end{figure}

The effective Zeno dynamics given by
\begin{align}
\L_Z[\rho]=-i[H_Z,\rho]+\gamma\big(\sigma^-\rho\sigma^+-\frac{1}{2}\{\sigma^+\sigma^-,\rho\}\big),\label{effLex}
\end{align}
is governed by the Zeno Hamiltonian $H_Z\,=\,\frac{\Delta}{2}\cos\theta\sigma_z$ and the incoherent part which stays unaffected. The relevant superoperators take the form
\begin{align}
\Q\ad_{H_m}^{-1}\Q[X]&=\frac{1}{\eta_1-\eta_2}\bigg(\bra{1}X\ket{2}\ket{1}\bra{2}-\bra{2}X\ket{1}\ket{2}\bra{1}\bigg),
\end{align}
and
\begin{align}
\D_1(t)[X]&=\P\L\bigg(f\Q_{12}\,+\,f^*\Q_{21}\bigg)\L\P[X],\\
\Q_{12}[X]&=\bra{1}X\ket{2}\ket{1}\bra{2},\\
f&=\frac{1-\exp\{ikt(\eta_1-\eta_2)\}}{ik(\eta_1-\eta_2)},
\end{align}
yielding the following results,
\begin{align}
\sfC&=\frac{1}{2}\Delta^2\sin^2\theta\,\left(\begin{array}{cc} 1&-1\\ -1& 1\end{array}\right),\\
\Omega&=|\Delta||\sin\theta|.
\end{align}
Fig.~\ref{fig:qubit} shows the relation between $\coh(H)$ and the bound $\Omega$ provided by our scheme. It illustrates in the simplest possible scenario the performance and limitations of our proposal. When  the effective Zeno Hamiltonian differs the most from the true undriven Hamiltonian ($\theta\,=\,\pi/2$) our methods provide the best bound ($\Omega\,=\,\coh(H)\,=\,1$). In all other cases, the bounds may be loose, to the extreme case of being trivially zero when the dynamics rests unaffected by the decoupling mechanism ($\theta=0,\,\pi$).

Most remarkably, this simple example shows that the coherence witness $\Omega$ is potentially tight.
\subsection{$N$-site spin chain with roller coaster energy landscape}
As a second example we take a $N$-site spin chain with nearest neighbor coupling $J$ and a roller coaster type energy landscape with a gap $E$ between the consecutive sites,
\begin{align}
	&H=\sum_n E_n\ket{n}\bra{n}+J\bigg(\sum_{n} \ket n\bra{n+1}+\ket{n+1}\bra{n}\bigg),\nonumber\\
	&E_n=E\times \big[n\mod 2\big].
\end{align}
The action of the environment is given by incoherent hoping among nearest-neighbor sites, described by jump operators $W_{nm}$
\begin{align}\label{eq:hopping}
&W_{nm}=\left\{\begin{array}{ll}
		\sqrt{\gamma} e^{\beta E }\ket{m}\bra{n}&\mathrm{~~if~}\mathrm{mod}_2n=1\,\&\,m=n\pm1,\\
		\sqrt{\gamma} e^{-\beta E }\ket{m}\bra{n}&\mathrm{~~if~}\mathrm{mod}_2n=0\,\&\,m=n\pm1,\\
		0&\mathrm{~~otherwise}.
	\end{array}\right.
\end{align}
The effective dynamics in the Zeno regime is given by
\begin{align}\label{eq:HZ}
	H_Z=\sum_i P_i H P_i.
\end{align}
Clearly, a more coarse grained decomposition in Zeno subspaces results in more off-diagonal terms from $H$ persistent in $H_Z$, whereas with less coarse grained Zeno subspaces (\emph{i.e.} more 'resolution' in the measurement setting) more off-diagonal terms are eliminated in the Zeno regime. Therefore the performance of $\Omega$ will strongly depend on the resolution of the measurement. It is worthy of mentioning that a coarse grained Zeno subspace decomposition $\{\PP_i\}$ leaves $\L_\phi$ unaffected if the fine-grained also does so, \emph{i.e.}, $\L_{\phi_\eff}=\L_{\phi}$, as is the case for the model considered here.

\begin{figure}
	\includegraphics[width=\linewidth]{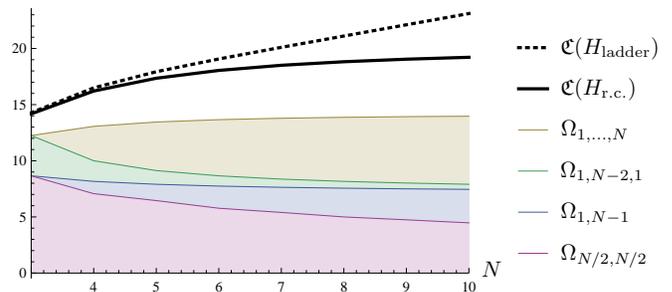}
\caption{Performance of the different measurement schemes for the rollercoaster [solid] and the ladder [dotted] models, at $J=5$. As these two Hamiltonians differ only in their diagonal entries, the values of $\Omega$ are identical for either of them. The value of $\coh(H)$ is not. This makes $\Omega$ a better or worse bound depending on each particular case. In addition, as explained in the text, the bounds become worse under coarse graining, being the $(1,N-2,1)$ and the $(1,N-1)$ measurements virtually equivalent for large number of sites. As discussed in Sec.~\ref{sec:universal_coh}, both $\coh(H)$ and $\Omega$ are independent of the type and strength of decoherence, and only determined by $H$ and the Zeno subspace decomposition.\label{fig:combined2D}.}
\end{figure}

Let us go back now to the discussion of the performance of witness $\Omega$. 
Fig.~\ref{fig:combined2D} shows $\coh(H)$ itself (dotted curve) and its lower bounds $\Omega_{d_1,\ldots,d_n}$, where subscript denotes the number of sites (dimension) in each subspace of the $\{\PP_i\}$ decomposition. We have evaluated $\Omega_{d_1,\ldots,d_n}$ for a variety of decompositions, and plot their performance at different number of sites $N=\sum_{i=1}^n d_i$. It is clear that the best bound is obtained with rank-1 (single-site) projectors. We can see as well, that for  increasing number of sites the coarse grained type of measurement, $\Omega_{1,N-2,1}$ approaches $\Omega_{1,N-1}$; which is to be expected.
More details on the behavior of $\Omega_{1,\ldots,N}$ can be seen in Fig.~\ref{fig:RC3D}: it shows $\coh(H)$ and $\Omega_{1,\ldots,N}$ obtained from a single-site resolving measurement, both as functions of the number of sites and coupling $J$. As can be seen $\Omega_{1,\ldots,N}$ follows the behavior of $\coh(H)$. 
\subsection{$N$-site spin chain with ladder energy landscape}
Here we take the $N$-site spin chain with the nearest neighbor coupling $J$ and a ladder type energy landscape characterized by the energy step $E$ 
\begin{align}
	&H=E\sum_n(N-n)\ket{n}\bra{n}+J\bigg(\sum_{n} \ket n\bra{n+1}+\ket{n+1}\bra{n}\bigg).
\end{align}
Similarly to the previous case, the action of the environment is given by incoherent hoping, Eq.~\eqref{eq:hopping}.
The effective dynamics in the Zeno regime takes analogous form as in the previous subsection, Eq.~\eqref{eq:HZ}, but with the energy levels decreasing linearly with $n$. As previously, the incoherent part of the dynamics is unaltered in the Zeno regime: $\L_{\phi_\eff}\,=\,\L_{\phi}$.
\begin{figure}
\includegraphics[width=\linewidth]{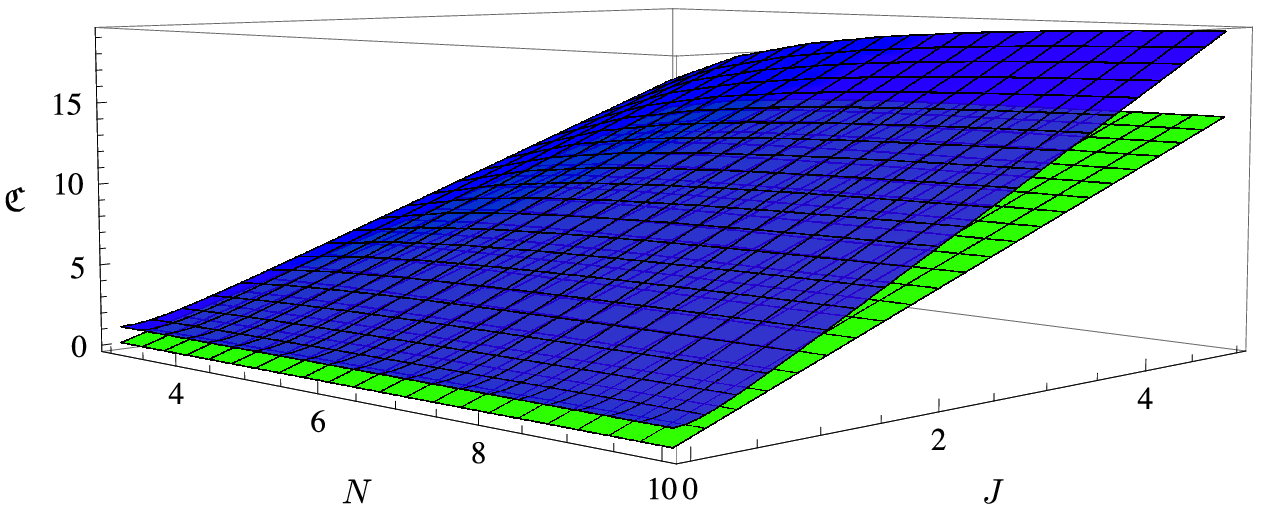}
\caption{[Upper] $\coh(H)$ for the rollercoaster Hamiltonian as a function of the number of sites N and the nearest-neighbor coherent hopping interaction $J$ varying from 0 to 5. Notice that for large number of sites, $\coh(H)$ becomes linear in $J$ and saturates at a constant value. [Lower] Single-site resolving measurement provides a good bound $\Omega_{1,\ldots,N}$ with similar behavior.\label{fig:RC3D}}

\vspace{.5cm}
\includegraphics[width=\linewidth]{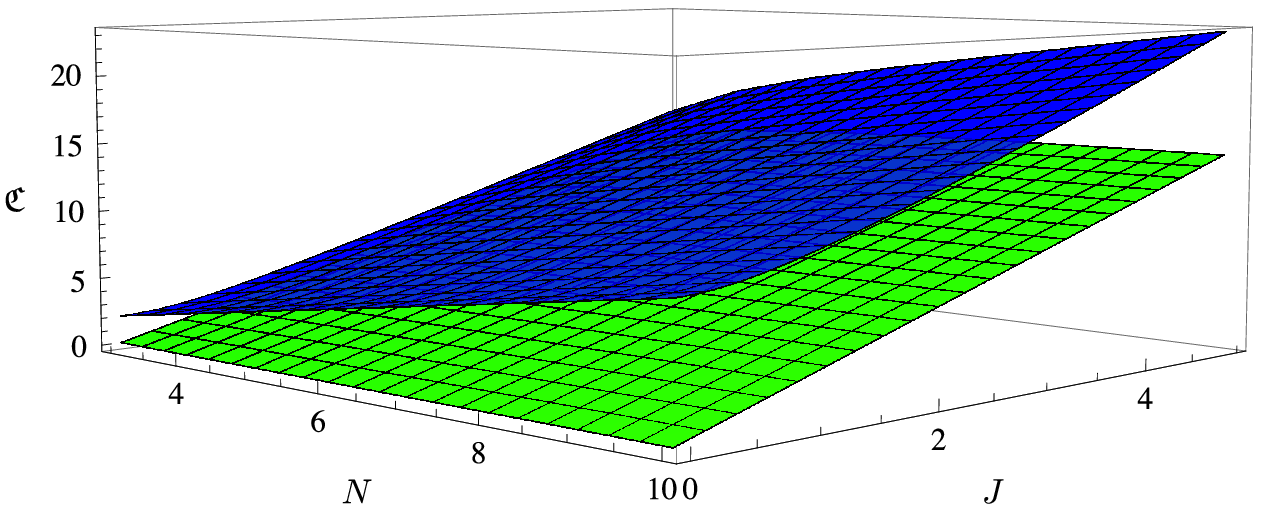}
\caption{\label{fig:ladder3D}[Upper] $\coh(H)$ for the ladder Hamiltonian as a function of the number of sites $N$, and nearest-neighbor coherent hopping interaction $J$ varying from 0 to 5. Notice that for large number of sites, $\coh(H)$ becomes linear in $J$ and $N$. [Lower] Single-site resolving measurement provides a lower bound $\Omega_{1,\ldots,N}$. The linearity in $J$ is captured by $\Omega_{1,\ldots,N}$, although not that in $N$.}
\end{figure}

Fig.~\ref{fig:combined2D} (dashed curve) shows results for $\coh(H)$ for the ladder model. The larger $N$, the looser are the bounds obtained. This is due to the fact that for a fixed value of $J$ and increasing $N$ the coherence measure $\coh(H)$ is dominated by the diagonal part of $H$ and is of order $N$. This is a general feature of our method: the more "aligned" with the eigenspaces of $H$ the Zeno subspaces are, the less effective our witness is. This occurs because in these situations the Zeno effect has essentially \emph{little coupling to suppress}.

Nevertheless, Fig.~\ref{fig:ladder3D} illustrates that even for the ladder model, our scheme provides bounds sensible to the strength $J$ and the measurement can reveal the magnitude of the couplings between the Zeno subspaces.

In both cases, roller coaster and ladder type energy landscape, one could consider also dephasing. It can be shown that also for this type of noise $\L_\phi\,=\,\L_{\eff}$. However, we have checked that in the examples provided dephasing does not bring anything new to the discussion and therefore we have focused on the incoherent hopping as a main source of decoherence.

\section{Discussion}\label{sec:discussion}

We have shown how a measurement protocol and analysis based on the notion of Zeno susceptibility can be used as a means for witnessing coherence. Our results constitute a proof-of-principle for using the Zeno effect, implemented by means of continuous driving, as a signature of nonclassicality. That the Quantum Zeno Effect is a genuinely quantum phenomenon is rather intuitive. The flipside of this statement is that dynamics susceptible to QZE are necessarily quantum. However, for noisy systems this susceptibility will only be partial. Hence, the amount by which the system's dynamics is affected upon continuous driving is an indicator of how much of said dynamics is due to coherent processes. Here we have shown how to make this statement quantitative, rigorous and operational. 

The approach outlined here may be extended towards situations in which some of our assumptions fail to hold. This could be done in a variety of ways, and an exhaustive study is well beyond the scope of this work. We  outline here a few of them.

\subsection{Generalizations}

The measurement protocol suggested has advantages and disadvantages: it does not rely on any specific choice of dimensions of $\PP_i$ but it relies on the fact that the system's Hilbert space can be decomposed into Zeno subspaces compatible with the noise (in accordance with Eq.~\eqref{eq:phiconditions}) and that this decomposition is known. This requirement may be lifted if one is willing to generalize our notion of coherence. We propose here candidates for extension of our approach. Given a Lindblad superoperator of the form Eq.~\eqref{eq:lindblad}, the Hamiltonian and jump operators are not uniquely defined. The transformations
\begin{subequations}
\begin{align}
	H&\rightarrow H +\frac{1}{2i}\sum_\mu\big(\alpha_\mu^* W_\mu- \alpha_\mu W_\mu^\dg\big)\\
	W_\mu&\rightarrow W_\mu+ \alpha_\mu \openone,\qquad\qquad \alpha_\mu\in\mathbb{C}
\end{align}
\end{subequations}
leave $\L$ invariant~\cite{tarasov_quantum_2008}. A universal measure of coherence which is independent of the particular representation of $\L$ could be given by
\begin{align}\label{eq:gencoh}
	\coh(\L)=\inf_{\alpha\in \mathbb{C}^n} \coh\left(H +\frac{1}{2i}\sum_\mu\big(\alpha_\mu^* W_\mu- \alpha_\mu W_\mu^\dg\big)\right),
\end{align}
where it is understood that $H$ and $W_\mu$ are the Hamiltonian and jump operators of any representation of $\L$. This quantity becomes zero if and only if the system's master equation can be entirely written in terms of jump operators --with no Hamiltonian--. In addition, it is invariant under the addition of a decoherence term --\emph{e.g.,} $\coh(\L)=\coh(\L+\L_\phi)$ for arbitrary $\phi$-- which supports the notion of being able to single out the coherent part of $\L$ despite the presence of noise. We leave as open problem to determine whether this quantity can be easily computed and/or measured. 

The lifting of $\coh(H)$ to $\coh(\L)$ can be seen as a relaxation or a tightening of our problem. On the one hand it tightens our framework by requiring to optimize a superoperator norm over all possible representations of $\L$. On the other hand, it is a relaxation because the only natural setup for using Eq.~\eqref{eq:gencoh} is when no assumptions are made about $W_\mu$'s. In particular, relaxation of assumptions \eqref{eq:phiconditions} seems a natural framework in which to consider quantities which depend solely on $\L$ and not one of its particular representations, as is Eq.~\eqref{eq:gencoh}.

Along with the general notion of coherence comes a natural question on a measure of decoherence that could be used as a reference.  In this work we refer to $\|\L\|$, the fastest timescales of the system's dynamics, as an upper bound to the fastest decoherence timescales (assuming that they are known) which dictate the smallness of $t$. 
However, a systematic approach is suggested by the so-called Leibnitz defect \cite{tarasov_quantum_2008} $\Delta_{\L}$,
\begin{align}
	\Delta_{\L}\big(A,B\big)\,=\,\L\big(AB\big)\,-\,\L\big(A\big)B\,-\,A\L\big(B\big),
\end{align}
that is zero if and only if $\L$ can be written solely in terms of a Hamiltonian. $\Delta_{\L}$ determines how much $\L$ fails to be of the commutator form $-i[H,\;]$. Thus, a suitable norm on $\Delta_{\L}$ is a reasonable candidate for a measure of the decoherence in $\L$.
Both measures will be discussed in more detail in future work.

\subsection{Implementations}
Experiments with continuous driving may be challenging in certain physical systems of interest. However, the QZE can be induced by other means, such as continuous measurement~\cite{facchi_quantum_2001} or dynamical decoupling~\cite{facchi_unification_2004}. Since the essence of our analysis relies not in the Zeno limit, but in how this limit is approached (the next-to-leading order contribution in the asymptotic expansions), it is not immediate that the same results will translate to these other setups. The ideas, however, should not vary in their essence. We leave the task of understanding how our ideas carry over to these setups for future work. 

In the case of implementation by means of continuous measurement, one may also track the set of measurement outcomes and not only the outcome of the final projective measurement. Then a parameter estimation could be applied, such as~\cite{gammelmark_bayesian_2013}, where authors discuss parameter inference from a continuously measured system with an application to quantum optics. The difficulty then relies in designing a general enough statistical parametric model capturing the specifics of our problem, while general enough to accommodate for unknown $\L$ dynamics. This procedure may allow to obtain more accurate estimates of $\Omega$ and even tighter bounds of~$\coh$.

Regarding the tightness of inequality \eqref{eq:mainresult}, one should note that only as much as is measured can be learnt: Hamiltonians that commute with the measurement Hamiltonian will be totally beyond reach of witness $\Omega$. In such cases, a different measurement basis and/or preparations will be required. For all other systems, $\Omega$ will provide a nontrivial lower bound to $\coh(H)$. As shown by examples, the more "noncommuting" the unknown Hamiltonian, the tighter inequality \eqref{eq:mainresult} will be. The virtue of this method is that it assumes noisy dynamics and, in our scheme, it provides bounds that are robust to the strength of noise.

Building the statistics needed for estimating $\dot\sfP$ is simple, but it may be challenging to work in the regime of small times and appropriate values of $k$. A deeper study of the timeframes for which the reduced system dynamics is Markovian (for large $k$) will allow to safely lift this small time constraint.

\section*{Acknowledgements}

The authors wish to thank early discussions with L. Heaney and A. Marais. Fruitful  discussions with S. Pascazio and M. Kastoryano are also acknowledged. We acknowledge support from the National Research Foundation and Ministry of Education, Singapore. A.M. also acknowledges  financial support from ERDF: European Regional Development Fund, the Spanish MICINN, through contract FIS2008-01236 and the ERC (Advanced Grant ``IRQUAT'').

\appendix
\section{Derivation of the effective dynamics}\label{app:TCL}

In the derivation of Eq.~\eqref{me} we absorb $k$ in the coupling Hamiltonian $H_m$, since our results will be independent of the strength of $H_m$.

We start decomposing the density operator $\rho$ in parts which we will call {\em populations} $\P[\rho]$ and {\em coherences} $\Q[\rho]$. We define the unitary superoperator $\U(t)=\exp it\ad_{H_m}$, which is just $\U(t)[x]=e^{itH_m}xe^{-itH_m}$, and write down the density operator in the interaction picture w.r.t. $H_m$, 
\begin{subequations}
\begin{align}
	\rho^I(t)&=\U(t)[\rho(t)]\\
	\L^I(t)&=\U(t)\L\, \U^*(t),
\end{align}
\end{subequations}
where we skip the composition symbol between superoperators and regard it as a product of operators in Hilbert-Schmidt space. The master equation Eq.~\eqref{eq:motherME} reads, in the interaction picture, 
\begin{align}
	\frac{d}{dt}\rho^I(t)=\L^I(t)[\rho^I(t)],
\end{align}
where, essentially, the Hamiltonian $H_m$ is canceled out. Our approach follows the derivation of the time-convolutionless master equation~\cite{breuer_theory_2002} which uses projection operator techniques~\cite{cohen-tannoudji_atom-photon_1992}, in order to derive the propagator from time $0$ to time $t$.

The time-evolution for the populations $\P[\rho^I(t)]$ and coherences $\Q[\rho^I(t)]$ then reads
\begin{subequations}
\begin{align}\label{eq:dP}
	\frac{d}{dt}\P[\rho^I(t)]&=(\P\L^I(t)\P+\P\L^I\Q)[\rho^I(t)],\\
	\frac{d}{dt}\Q[\rho^I(t)]&=\Q\L^I(t)[\rho^I(t)].
\end{align}
\end{subequations}
Now, let $\Gamma(t,t_0)=\mathbb{T}\exp\int_{t_0}^tds\Q\L^I(s)$ be the time ordered exponential (see~\cite{breuer_theory_2002}), {\em e.g.} the solution to
\begin{align}
	\frac{d}{dt}\Gamma(t,s)=\Q\L^I(t)\Gamma(t,s),
\end{align}
with boundary condition $\Gamma(t,t)=\id$. With this one can solve for the coherences $\Q[\rho^I(t)]$  
\begin{align}\nonumber
	\Q[\rho^I(t)]=\,&\Gamma(t,t_0)\Q[\rho^I(t)]\\
	&+\int_{t_0}^tds\,\Gamma(t,s)\Q\L^I(s)\P[\rho^I(s)],
\end{align}
which used in Eq.~\eqref{eq:dP} yields
\begin{align}\label{ref-tcl}
	\frac{d}{dt}\P[&\rho^I(t)]=\\
\nonumber
		&\P\L^I(t)\left(\P[\rho^I(t)]+\int_0^tds\,\Gamma(t,s)\Q\L^I(s)\P[\rho^I(s)]\right).
\end{align}

We are interested in writing the time-derivative of $\P[\rho(t)]$ at time $t$ as a function of the preparation $\P[\rho(0)]$, therefore we can write the time-evolution channel (in the interaction picture) and the density operator as
\begin{align}
	\E(t,t_0)=\mathbb{T}\exp \int_{t_0}^t ds \L^I(s),\qquad \rho^I(t)=\E(t,s)[\rho^I(s)].
\end{align}
Then, the evolution of populations is given by
\begin{align}
\frac{d}{dt}\P[\rho^I(t)]=&\,\P\L^I(t)\bigg(\P\E(t,0)\,+\nonumber\\
 &+\,\int_0^tds\,\Gamma(t,s)\Q\L^I(s)\P\E(s,0)\bigg)[\rho^I(0)].
\end{align}
Writing in the Schroedinger picture we obtain
\begin{align}\label{mequation}
	\frac{d}{dt} &\P[\rho(t)]=\P\L\P\E(t,0)[\rho(0)]\\
\nonumber
	&+\P\L\int_0^tds\,\U^*(t)\Gamma(t,s)\U(s)\,\Q\L\P\E(s,0)[\rho(0)]
\end{align}
Several considerations are worth making about this expression. The first line can be conveniently expressed as $\L_\eff[\rho(t)]$ , where $\L_\eff=\P\L\P$. As we will see, this represents the dominant part of the dynamics at time $t$. On the other hand, the essential part of the second line is characterized by the memory kernel
\begin{align}
	\K(t)=\int_0^tds\,\U^*(t)\Gamma(t,s)\U(s)\, \Q\L\P\E(s,0)
\end{align}
This operator characterizes the nonmarkovianity of $\P[\rho(t)]$, as its evolution cannot be uniquely determined without reference to the coherences $\Q[\rho]$. The operator $\K(t,s)$ provides the accumulated non-Markovianity at time $t$. Numerical evidence shows that for relatively large values of $k\sim \|\ad_{H_m}\|$, $\K(t)$ is of order $1/k$ up to some time $T$, after which it becomes an important contribution, suggesting that after time $T$ the system acquires enough memory to make the non-Markovian effects relevant. A detailed study of this phenomenon is beyond the scope of this work. 

Nevertheless, it is interesting to note that the contribution of the second line in Eq.~\eqref{mequation} characterizes the adiabatically eliminated transitions among Zeno subspaces, mediated by coherences originated at time $s$, and evolving over time until $t$. The overall effect of this is a correction to the leading order dynamics. The relative relevance of this term will dictate whether transitions among subspaces occur due to incoherent first-order processes or through these virtual second-order transitions.

Instead of attempting a full solution of Eq.~\eqref{mequation} we find it constructive to consider the evolution at small times $t$ such that $\epsilon=t\|\L\|\ll1$. In this case, $\|\L\|$ has the dimension $1/t$ and determines the magnitude of the fastest timescales arising in Eq.~\eqref{meq}. With the small parameter introduced, we can write 
\begin{align}
&\E(t,0)\,=\,\id+\int_0^tds\L^I(s)+O(\epsilon^2),\\
&\Gamma(t,s)\,=\,\id+\int_s^tds'\Q\L^I(s')+O(\epsilon^2).
\end{align}
We use the above expansions to write the equation for the evolution of the populations 
\begin{align}
\frac{d}{dt}&\P[\rho(t)]=\\
\nonumber&=\P\L\left(\P+\int_0^tds \,\U(s-t)\L\P+\mathcal{O}(\epsilon^2)\right)[\rho(0)]
\end{align}%
Here we note that the action of unitaries (related to 'adiabatic elimination term') can be represented as
\begin{align}
\int_0^tds \,\U(s-t)\,=\,t\P\,+\,\Q\frac{\id-e^{-it\ad_{H_m}}}{i\ad_{H_m}}\Q.
\end{align}
\begin{widetext}
Therefore, we can write the second order term as
\begin{align}
\P\L \int_0^t&ds \,\U(s-t)\L\P=t\P\L\P\L\P+\P\L\Q\frac{\id-e^{-it\ad_{H_m}}}{i\ad_{H_m}}\Q\L\P,
\end{align}
and finally
\begin{align}
&\frac{d}{dt}\P[\rho(t)]=\P\L\bigg(\P(\id+t\P\L\P)+\Q\frac{\id-e^{-it\ad_{H_m}}}{i\ad_{H_m}}\Q\L\P+O(\epsilon^2)\bigg)[\rho(0)]).
\end{align}

One can check that chosing $k\sim\|\ad_{H_m}\|$ of the order $1/t$ so that $t\ad_{H_m}\sim 1$ leads to superoperators $\D_0$ and $\D_1$ of similar magnitudes. This is the desirable regime to work in, which will render the dependency of $\D_1$  in $k$ most visible despite statistical and experimental errors.

\end{widetext}%

\section{Proofs of results regarding superoperators}\label{app:bounds}
\newcommand{\ac}{\mathrm{ac}}
We provide here proofs of some of the facts stated in the text that may not be obvious to all readers.

\subsection{Implications of the preferred basis}\label{app:phidrops}
Introducing the anticommutator superoperator
\begin{align}
	\ac_A[X]=\frac{1}{2}(A X+X A)
\end{align}
one can check that for selfadjoint $A=A^\dg$, $\ac_A$ is also self-adjoint,
\begin{align}
	\ac_A^*=\ac_A.
\end{align}
Using this notation, $\L_\phi$ reads
\begin{align}
	\L_\phi=\phi-\ac_{\phi^*({\small \openone})}
\end{align}
and our assumptions~\eqref{eq:phiconditions} read
\begin{subequations}
\begin{align}
\label{Qphi}
	\Q \phi [P_i]&=0\\
\label{Qphidg}
	\Q\phi^*[P_i]&=0.
\end{align}
\end{subequations}
From this, we go on to show that 
\begin{subequations}
\begin{align}
\label{eq:QLP}
	\Q\L_\phi[P_i]&=0\\
\label{eq:PLQ}
	(\L_\phi\Q)^*[P_i]&=0.
\end{align}
\end{subequations}
From Eq.~\eqref{Qphidg} it is readily seen that 
\begin{align}
	\Q\ac_{\phi^*(\openone)}[P_i]=0
\end{align}
which shows that $\Q\L_\phi[P_i]=\Q \phi[P_i]-\Q\ac_{\phi^*(\openone)}[P_i]=0$, Eq.~\eqref{eq:QLP}. Using self-adjointness of $\Q$ and $\ac_{\phi^*(\openone)}$ we have that Eq.~\eqref{eq:PLQ} reads
\begin{align}
\nonumber
	\Q \L_\phi^*[P_i]&=\Q(\phi^*-\ac_{\phi^*(\openone)})[P_i]\\
	&=0.
\end{align}
%
%, we show how Eqs.~\eqref{eq:PLQzero} follows. Notice that $\L_\phi=\phi-\ac_\phi$, where
%
%%This superoperator is self-adjoint, $\ac_\phi^*=\ac_\phi$ and it is clear that
%%\begin{align}
%%	\Q\ac_\phi\P[X]=\sum_{r\neq s}\sum_iP_r (\phi^*[\openone]P_iXP_i+P_iXP_i\phi^*[\openone])P_s
%%\end{align}
%
%Obviously $\Q\L_{\phi}\P[P_i]=\Q\L_{\phi}[P_i]$ and $\Q\L_{\phi^*}\P[P_i]=\Q\L_{\phi^*}[P_i]$. Let us write explicitly the former one
%
%\begin{align}
%\Q\L_\phi[P_i]&=\sum_{r\neq s}P_r\phi(P_i)P_s-\frac{1}{2}\sum_{r\neq s}P_r\big(\phi^*(\openone)P_i+P_i\phi^*(\openone)\big)P_s\,\nonumber\\
%&=-\frac{1}{2}\big(\sum_{r\neq s}P_r\phi^*(\openone)P_s\delta_{is}+\sum_{r\neq s}\delta_{ir}P_r\phi^*(\openone)P_s\big)\nonumber\\
%&=-\frac{1}{2}\sum_{r\neq s}\delta_{ik}\big(P_r\phi^*(\openone)P_s+P_s\phi^*(\openone)P_r\big)\nonumber\\
%&= 0,
%\end{align}
%where we used $\Q\phi(P_n)=\Q\phi^*(P_n)=0$ and $\openone=\sum_jP_j$. $\Q\L_{\phi^*}\P[P_i]=0$ can be shown in analogous fashion.

This implies that Eq.~\eqref{eq:Cmu}, $[\sfT_\mu]_{ij}=-\tr[\tilde P_i\, \L\Q_\mu\L[\tilde P_i]]$ when written as
\begin{align}
	[\sfT_\mu]_{ij}=-\frac{1}{\sqrt{d_id_j}}\tr[(\Q\L^*[P_i])^\dagger\, \Q_\mu\, \Q\L[P_j]] 
\end{align}
can be expressed as $[\sfT_\mu]_{ij}=\tr[B^\dg\Q_\mu[A]]$, with 
\begin{subequations}
\begin{align}
	A&=\Q(-i\,\ad_H+\L_\phi)[P_j]=-i\,\ad_H[P_j]\\
	B&=\Q(-i\,\ad_H+\L_\phi)^*[P_i]=i\,\ad_H[P_i]
\end{align}
\end{subequations}
so that $[\sfT_\mu]_{ij}$ reduces to
\begin{align}
\nonumber
	[\sfT_\mu]_{ij}&=\tr[(\ad_H[\tilde P_i])^\dagger\, \Q_\mu\, \ad_H[\tilde P_j]]\\
	&=\tr[\tilde P_i\,\ad_H\Q_\mu \ad_H[\tilde P_j]].
\end{align}
%{\color{red}This - is here something missing?? } 
%
%
%
%{\color{red} I understand that the part below is the alternative proof.}
%For a case when we act on a density operator $\chi=\sum_jr_jP_j$ we get 
%
%\begin{align}
%\Q\L_\phi[\chi]&=\sum_{r\neq s}P_r\phi(\chi)P_s\,-\,\frac{1}{2}\sum_{r\neq s}\big(\phi^*(\openone)\chi+\chi\phi^*(\openone)\big)P_s\,\nonumber\\
%&=\sum_jr_j\sum_{r\neq s}P_r\phi(P_j)P_s\,+\nonumber\\
%&-\frac{1}{2}\sum_jr_j\sum_{r\neq s}P_r\big(\phi^*(\openone)P_j+P_j\phi^*(\openone)\big)P_s\,\nonumber\\
%&=-\frac{1}{2}\sum_j\big(r_s\sum_{r\neq s}P_r\phi^*(\openone)P_s+r_r\sum_{r\neq s}P_r\phi^*(\openone)P_s\big)\nonumber\\
%&= 0,
%\end{align}

\subsection{Monotonicity of the operator norm}\label{app:monotonicity}

Recall that $\{\tilde P_i\}$ is an orthonormal basis in a subspace $V$ of $L(\mathcal H)$ and $\{\tilde P_i, T_r\}$ is its extension to $L(\mathcal H)$. In this sense $\sfC_{ij}=\tr[\tilde P_i \ad_H^2 [\tilde P_j]]$ is the matrix representation of superoperator $\ad_H^2$ restricted to subspace $V$. It is positive semidefinite as can be seen by 
\begin{align}
	x^\top \sfC \,x=\tr\big[\ad_H[X]^\dagger\, \ad_H[X]\big]\geq0,
\end{align}
where $X=\sum_i x_i \tilde P_i$. Thus, $\sqrt{\sfC}$ is uniquely defined and its operator norm is given by
\begin{align}
\nonumber
	\Big\|\sqrt\sfC\Big\|_{\infty}&=\sup_{\|x\|=1}\Big\|\sqrt\sfC x\Big\|\\
\nonumber
	&=\sup_{\|x\|=1}x^\top \sfC x\\
\label{eq:operatorboundnorm}
		&=\sup_{\|x\|=1}\Big\| \ad_H \big[\sum_j x_j \tilde P_j\big]\Big\|_2,
\end{align}
where $\|x\|=\sqrt{x^\top x}$ is the standard Euclidean norm and $\|X\|_2$ is the Hilbert-Schmidt norm $\|X\|_2=\sqrt{\tr[X^\dagger X]}$. For any $x$ such that $\|x\|=1$ we have that $X=\sum_i x_i \tilde P_i$ is
\begin{align}
	\|X\|_2 =\Big(\sum_{ij}x_i^* x_j \tr[\tilde P_i \tilde P_j]\Big)^{1/2}=1.
\end{align}
Hence, we can upper bound Eq.~\eqref{eq:operatorboundnorm} by relaxing the maximization to all operators in $L(\mathcal H)$ normalized w.r.t the 2-norm, thus obtaining the induced Hilbert-Schmidt superoperator norm
\begin{align}
	\|\sqrt\sfC\|_\infty \leq \sup_{\|X\|_2=1}\big\| \ad_H[X]\big\|_2\equiv\|\ad_H\|_2.
\end{align}

\subsection{Computation of the induced Hilbert-Schmidt norm}\label{app:opnorm}
Here we show that $\|\ad_H\|_2=\lambda_{\max_{}}(H)-\lambda_{\min_{}}(H)$, where $\lambda_i(H)$ are $H$'s eigenvalues and we will drop their dependence on $H$. Notice that eigenvectors of $\ad_H$ are given by $\ket{\psi_\alpha}\bra{\psi_\beta}$, where $\ket{\psi_\alpha}$ constitute the eigenbasis of $H$. Then, any operator $X$ can be expressed in the eigenbasis of $\ad_H$, 
\begin{align}
	X=\sum_{\alpha\beta}x_{\alpha\beta}\ket{\psi_\alpha}\bra{\psi_\beta}
\end{align}
and hence 
\begin{align}
\nonumber
	\|\ad_H(X)\|_2&=\|\sum_{\alpha\beta}x_{\alpha\beta}(\lambda_\alpha-\lambda_\beta)\ket{\psi_\alpha}\bra{\psi_\beta}\|_2\\
\nonumber
	&=\left(\sum_{\alpha\beta}|x_{\alpha\beta}|^2(\lambda_\alpha-\lambda_\beta)^2\right)^{1/2}\\
	&\leq\lambda_{\max_{}}-\lambda_{\min_{}}
\end{align}
%On the other hand, the constraint $\|X\|_2=1$ reads $\sum_{\alpha\beta} x_{\alpha\beta}=1$. Hence  $\|\ad_H\|_2^2$ is a convex combination of the squared gaps $(\lambda_\alpha-\lambda_\beta)^2$, maximized $X=\frac{1}{2}(\ket{\psi_{\max_{}}}\bra{\psi_{\min_{}}}+\ket{\psi_{\min_{}}}\bra{\psi_{\max_{}}})$ 
%\begin{align} 
%	&\leq |E_{\max_{}}-E_{\min_{}}|.
%\end{align}
Clearly, $\sum_{\alpha\beta}|x_{\alpha\beta}|^2=1~\Leftrightarrow~\|X\|_2=1$, thus the bound is attainable, with $X=\ket{\psi_{\max_{}}}\bra{\psi_{\min_{}}}\in L(\mathcal H)$, where $\ket{\psi_{\max_{}}},\ket{\psi_{\min_{}}}$ are eigenvectors corresponding to $\lambda_{\max_{}}$, $\lambda_{\min_{}}$ resp. Thus
\begin{align}
\nonumber
	\|\ad_H\|_2&=\sup_{\|X\|_2=1}\big\| \ad_H[X]\big\|_2\\
	&=\lambda_{\max_{}}-\lambda_{\min_{}}.
\end{align}

\bibliographystyle{apsrev}

\end{document}